# Sharing-Based Channel Access Procedure For Next Generation of Wireless LAN

Qing Xia, *Member, IEEE* and Salvatore Talarico, *Member, IEEE*

Sony Corporation of America, R&D US Laboratory, 1730 N First Street, San Jose, CA 95112, USA.
E-mail: {Qing.Xia, Salvatore.Talarico}@sony.com

*Abstract*—This paper proposes a new channel access procedure to mitigate the channel access contention in next generation of Wireless Local-Area Networks (WLANs) by allowing cooperation among devices belonging to same network, while maintaining high flexibility in terms of how each device may contend the medium. After introducing the details of the proposed procedure, which is here referred to as sharing-based protocol, an analytical analysis is provided to compare it with the two state-of-art protocols currently adopted in IEEE 802.11 standard, i.e, Enhanced Distributed Channel Access (EDCA)-based and trigger-based protocol. In this regards, closed form expressions are derived to evaluate the success probability of channel access for each protocol. In order to show the merit of the proposed procedure, a comprehensive system level analysis is also provided, which highlights that the proposed procedure outperforms the two state-of-art protocols in terms of mitigating the End-to-End (E2E) delay and allowing a better spectrum utilization by reducing the overall congestion in the system.

## I. INTRODUCTION

The Wi-Fi technology [1], [2] is one of the major technological achievement of the past 25 years, which has profoundly impacted society and has now become embedded in it by being an essential component for our daily lives. Currently, the latest Wi-Fi products in the market are using Wi-Fi 6(E) release, while from standardization perspective the IEEE working group devoted to evolving specifications for WLAN is working toward the end of the Wi-Fi 7 [3]–[5]. Wi-Fi 7 is meant to enhance prior releases by enabling high capacity applications for immersive user experience such as AR/VR and social gaming, as well as industrial applications that require real-time capabilities like factory monitoring that all rely on Wi-Fi to deliver high data throughput and low latency. These applications offer a wide range of business solutions that have the potential to revolutionize both personal and enterprise activities in the next decade and through a set of advanced features would allow to achieve multi-Gbps throughput, sub-10 ms latency and packet losses lower than 0.1%.

In order to mitigate the E2E latency of such Wi-Fi system, multiple components must be taken into account, including the propagation, the processing, the queuing, the channel access, and re-transmission delay. The propagation delay is the smallest of all of them, and is fixed as it depends on the speed of light. As for the processing delay, this is purely implementation and hardware specific, and highly application dependent. Among the remaining three components that characterize the E2E delay, the channel access delay is the most predominant one and it also impacts intrinsically the queuing delay and the re-transmission delay. A proper channel access protocol not only allows a better utilization of the spectrum, and reduces the channel access delay of a Wi-Fi system, but may also ensure that inter-Radio Access Technology (RAT) and intra-RAT technologies may co-exist well with each other [6]. Currently, Wi-Fi systems support two types of channel access protocols [7]: i) EDCA-based channel access protocol, and ii) trigger-based channel access protocol. In the EDCA-based channel access protocol, each non-Access Point (AP) Station (STA) senses the medium status and maintains a Back-Off (BO) timer. The BO timer value is randomly selected based on the Access Category (AC) used, which depends of the QoS requirements, and a non-AP STA starts counting down the BO each time the medium is assessed to be idle and it gets paused otherwise. A non-AP STA initiates a transmission in the medium when the BO timer reaches zero. The EDCA-based channel access protocol allows a non-AP STA to contend the channel at any time with no restrictions. However, this high level of flexibility impacts the overall E2E delay and user experience, since collisions across non-AP STAs may more likely occur as every device contends the medium with each other [8]. In the trigger-based channel access protocol, the AP works as a centralized coordinator/scheduler, which first collects information (e.g., buffer status) from the associated non-AP STAs and then solicits Uplink (UL) transmissions through a basic trigger frame. The trigger-based channel access protocol allows to reduce congestion across non-AP STAs through the centralized scheduling. However, this advantage comes with decreased channel access flexibility [9], [10].

In this paper, a *sharing-based channel access protocol* is proposed, which on one hand effectively reduces the congestion across devices, but on the other hand maintains high flexibility in contending the channel. The sharing-based protocol is based upon the principle that non-AP STAs cooperate with each other and that even though any non-AP STA may access the channel access, they may share its Transmission Opportunity (TxOP) with other STAs (including the AP) with the benefit that these devices would not need to content the channel with the STAs sharing its TxOP.

The remaining of this paper is organized as follows. Sec. II provides a summary of the related work. In Sec. III, the new sharing-based channel access protocol is described and detailed comparison in terms of typical transmission timeline with the EDCA-based and trigger-based channel access protocol is provided. In Sec. IV, closed form expressions to compare the three protocols in terms of success probability of channel access are derived. Sec. V provides a comprehensive comparison of the three protocols through system level evaluations. Finally, the paper is concluded in Sec. VI.

## II. RELATED WORK

Extensive research has been performed in the past years with the aim to improve the random channel access performance. In this matter, new distributed channel access algorithms have been proposed for densely deployed networks to account specifically for their load conditions. For example, authors in [11] proposed to assign different initial BO values to each STA according to the network load, while in [12] the authors proposed a binary exponential BO algorithm based on the severity of the collisions occurring under certain packet generation patterns. Some researchers have also focused on the coexistence between centralized and distributed channel access mechanisms, and they have proposed several novel channel access algorithms. In [9], the authors proposed a trigger-based asynchronous CSMA/CA mechanism that enables both the multiple-packet reception dimension and the time dimension to adapt the channel access probabilities so that to allow STAs to start transmissions in an already busy channel as long as the current usage level is below a certain threshold. In [10], the authors proposed a mechanism to enable the STAs to dynamically adapt their channel access priorities in order to mitigate contention. In [13], the authors introduced a new multiple-users (MU)-EDCA channel access, which temporarily de-prioritizes the medium access for STAs to allow the associated AP to access the channel with higher likelihood. An Orthogonal Frequency-Division Multiple Access (OFDMA)-based hybrid channel access method is proposed in [14], which tunes the BO window to help mitigating collisions among STAs. In [15], the authors proposed an enhanced MU-EDCA algorithm meant to guarantee the channel access of the AP by stopping the associated STAs from contending the channel. In addition, researchers have proposed many mechanisms to share a TxOP between APs, and as an example in [16] the authors designed two TxOP sharing strategies between coordinated APs, where one employs coordinated Time Division Multiple Access (TDMA) and the other coordinated Spatial Reuse (SR).

With that said, based on the best of our knowledge, there is no work reported in the literature which proposes any TxOP sharing protocols initiated by the non-AP STA.

## III. SHARING-BASED PROTOCOL

The novel sharing-based protocol is designed with the main aim to mitigate the contention among non-AP STAs by allowing a level of cooperation among them. In particular, the proposed protocol is based upon the logic that when a non-AP STA obtains a TxOP, that we refer here as the TxOP holder, its TxOP could be in part shared with other non-AP STAs, that we refer here as the shared non-AP STAs. In this matter, the proposed protocol allows the shared non-AP STAs to more efficiently transmit and utilize the channel during the shared TxOP without contending for channel among them. Thus, the sharing-based protocol helps reducing the contention among non-AP STAs while maintaining the flexibility of channel access for the individual non-AP STA, since they are still allowed to contend independently the channel once there is no active TxOP where they either serve as TxOP holder or shared non-AP STA.

For a non-AP STA to initiate a shared TxOP, the TxOP holder sends first an $RTS^s$ frame to the associated AP with indication of both its willingness of share the TxOP and other sharing information, such as the shared TxOP duration allocated to each of the shared non-AP STAs participating in the shared TxOP and the traffic priority requirements. At this point, the associated AP of the TxOP holder assists the coordination among the TxOP holder and shared non-AP STAs participating in the shared TxOP. More specifically, after receiving the $RTS^s$ frame from the TxOP holder, the AP loops through the identified shared non-AP STAs participating in the shared TxOP by polling each of them through a $CTS^s$ frame, which carries the sharing information as indicated in the $RTS^s$ frame. Upon receiving the $CTS^s$ frame, the shared non-AP STAs polled by its associated AP could transmit $Data$ frames without contending the channel access within the allocated shared TxOP duration of the shared TxOP.

A shared non-AP STA may send a $CTS_{rej}$ frame to the associated AP if any of the following conditions applies:

- the shared non-AP STA finishes transmission earlier than the assigned shared TxOP duration;
- the non-AP STA approaches the end point of the assigned shared TxOP duration;
- the non-AP STA has no data to deliver.

The AP of the TxOP holder may either keep polling the shared non-AP STAs participating in the shared TxOP until it meets the TxOP limitation or polls only a set of the shared non-AP STAs and terminates the shared TxOP earlier than the TxOP limitation if no longer receiving any data from the shared non-AP STAs. In both cases, the AP of the TxOP holder will broadcast a $CF_{end}$ frame to clear the Network Allocation Vector (NAV).

### A. Comparison with Existing Protocols

Fig. 1a, Fig. 1b and Fig.1c show typical transmission timelines for the EDCA-based, the trigger-based and also for the proposed sharing-based protocol, respectively. In each timeline, the frame transmissions are from 5 nodes, including one AP, which is denoted as node 1, and four associated non-AP STAs, which are marked as node 2 - 5, respectively.

Fig. 1a shows the timeline for the EDCA-based protocol, and each non-AP STA contends the channel independently without any cooperation with other STAs. Fig. 1b shows the timeline for the trigger-based protocol. In this case, the AP contends for channel access and once it has obtained the TxOP it sends the Buffer Status Report Poll (BSRP) frame, $BSRP$, to solicit the Buffer Status Report (BSR) frames, $BSR$, from the polled non-AP STAs. Then, based on the obtained BSR information, the AP transmits a Basic Trigger (BT) frame, $BT$, to trigger the UL MU-Data frames from the triggered STAs. Lastly, the AP sends a Block Ack (BA) frame, $BA$, as a response to the reception of MU-Data. Finally, Fig. 1c illustrates the timeline for the sharing-based protocol, where in this example node 5 acts as a TxOP holder.

## IV. STATISTICAL CHANNEL ACCESS MODEL

Consider a network composed by two adjacent Basic Service Sets (BSSs): one is the reference BSS where the intended transmissions occur, and the other is an Overlapping Basic Service Set (OBSS). Assume that all BSSs have the same coverage area as the corresponding APs transmit with the same transmitting power. Each BSS is modelled by a circular region with radius $r$ where the corresponding AP is located in the center, and its coverage area is equal to $A = \pi r^2$. Each AP is associated with $M$ non-AP STAs, which are distributed within the area $A$ according to a Poisson Point Process (PPP) with intensity $\lambda = M/A$, where the probability that $m$ non-AP STAs would be located in the coverage area $A$ is

$$P_A[m] = \frac{(\lambda A)^m}{m!} e^{-\lambda A}. \quad (1)$$

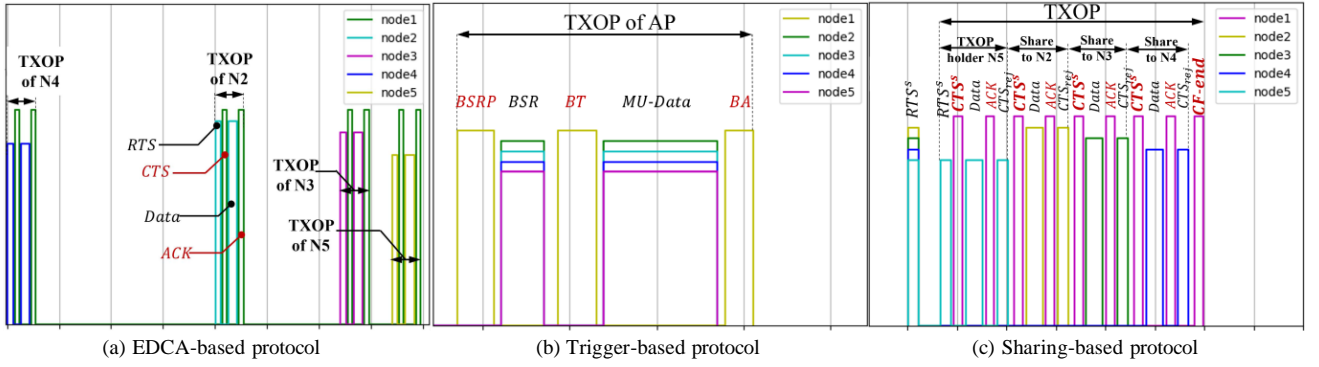

Fig. 1: Illustration of the EDCA, trigger-based and new sharing-based protocol.

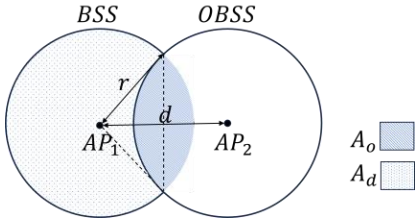

Fig. 2: Illustration of the network topology.

Assume that the reference BSS and the OBSS are located so that their coverage area would be overlapping and the respective APs would be distanced by $d$, where $d \in [0, 2r]$, as illustrated in Fig. 2. Let $A_o$ denote the overlapped coverage area between the two BSSs. Let us assume that the power received by any non-AP STA located within $A_o$ from the unintended AP is sufficient to prevent those non-AP STAs from accessing the channel as that transmission would cause interference, while an adjacent AP may not cause interference if a non-AP STA is located outside $A_o$, and let define this area as $A_d$. With that said, the overlapped coverage area $A_o$ is

$$A_o = 2\left(r^2 \cos^{-1}\left(\frac{d}{2r}\right) - \frac{d}{4}\sqrt{4r^2 - d^2}\right), \quad (2)$$

while $A_d = A - A_o$.

### A. EDCA-based Channel Access

Assume that each non-AP STA associated with the reference BSS, which are distributed within $A_d$ according with a PPP with intensity $\lambda_A$, generates a new frame at a rate given by $\lambda_{A_d} = 1/(a\tau)$, where $a$ is a constant indicating the inverse of the activity rate, and $\tau$ is the frame duration. In this case, the aggregated rate generated by $i$ non-AP STAs associated with the reference BSS, where all transmissions are orthogonal in time domain, is $\lambda'_{A_d} = i\lambda_{A_d}$. With that said, the probability that $n$ out of $i$ devices are active during a vulnerable window of $2\tau$, meaning that $n$ devices perform transmissions which overlap in time domain across them and cause intra-BSS collisions to each other, is given by

$$P_{A_d}[n|2\tau] = \frac{(2\tau\lambda'_{A_d})^n}{n!} e^{-2\tau\lambda'_{A_d}}. \quad (3)$$

Then, the probability that a non-AP STA associated with the reference BSS experiences intra-BSS collisions in $A_d$ is

$$P^{fail}_{A_d} = \sum_{i=1}^{\infty} P_{A_d}[i]\left(1 - P_{A_d}[n=0|2\tau]\right)$$
$$= \sum_{i=1}^{\infty} \frac{(\lambda_A A_d)^i}{i!} e^{-\lambda_A A_d}\left(1 - e^{-\frac{2i}{a}}\right). \quad (4)$$

Conversely, assume that each of the non-AP STAs associated to both the reference BSS or the OBSS, which are distributed within $A_o$ according with a PPP with intensity $\lambda_A$ and $\lambda_B$, respectively, generates a new frame at a rate given by $\lambda_{A_o} = 1/\beta\tau$, where $\beta$ is a constant indicating the inverse of the activity rate. In this case, the aggregated traffic generated by $j$ non-AP STAs associated with the OBSS, where all transmissions are orthogonal in time domain, is $\lambda'_{A_o} = j\lambda_{A_o}$. With that said, the probability that $n$ out of $i + j$ devices are active during a vulnerable window of $2\tau$, meaning that $n$ devices perform transmissions which overlap in time domain across them and cause both inter and intra-BSS collisions to each other, is given by

$$P_{A_o}[n|2\tau] = \frac{(2\tau\lambda_{A'})^n}{n!} e^{-2\tau\lambda_{A'}}, \quad (5)$$

where $\lambda_{A'} = \lambda'_{A_d} + \lambda'_{A_o}$.

Then, the probability that a non-AP STA associated with the reference BSS experiences both intra and inter-BSS collisions in $A_o$ is

$$P^{fail}_{A_o} = \sum_{i,j=1}^{\infty} P_{A_o}[i] P_{A_o}[j]\left(1 - P_{A_d}[n=0|2\tau]\right)$$
$$= \sum_{i,j=1}^{\infty} \frac{(\lambda_A A_o)^i (\lambda_B A_o)^j}{i! j!} e^{-A_o \lambda'}\left(1 - e^{-2\left(\frac{i}{a} + \frac{j}{\beta}\right)}\right), \quad (6)$$

where $\lambda = \lambda_A + \lambda_B$.

Finally, the probability that non-AP STAs within the coverage area of a reference BSS will be able to access the channel and transmit, also called success probablity of channel access, is given by:

$$P_{success} = 1 - \left(\frac{A_d}{A} P^{fail}_{A_d} + \frac{A_o}{A} P^{fail}_{A_o}\right). \quad (7)$$

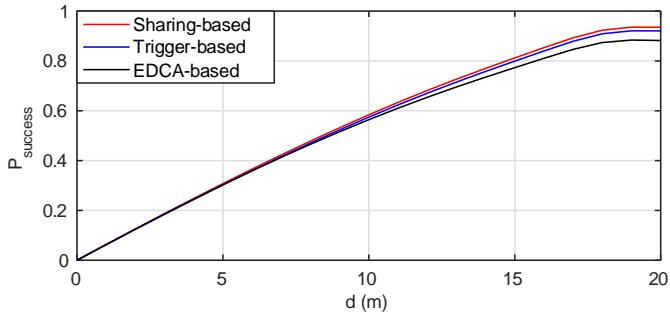

Fig. 3: Success probability of channel access as function of the distance between BSSs.

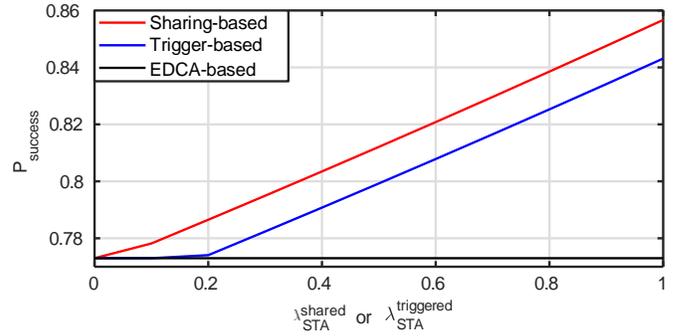

Fig. 4: Success probability of channel access as function of the ratio of participating non-AP STAs within a shared TxOP.

*B. Trigger-based Channel Access*

For the trigger-based channel access protocol, the formula to calculate $P_{success}$ is the same as that in Sec. IV-A, except for (3) and (5), where $\lambda'_{A_d}$ is replaced with $\lambda^{trigger}_{A_d}$ which is defined as follows

$$\lambda^{trigger}_{A_d} = \lambda'_{A_d}\left(1 - \lambda^{triggered}_{STA}\right) + \lambda^{trigger}_{g}, \quad (8)$$

where $\lambda^{triggered}_{STA}$ is the set of triggered non-AP STAs, which transmit uplink transmission upon receiving a trigger frame from the associated AP, thus they don't contend the channel, and $\lambda^{trigger}_{g} = 1/(\omega\tau)$ represents the trigger frame generation rate from the associated AP, where $\omega$ is a constant value indicating the frequency of the trigger frame.

*C. Sharing-based Channel Access*

For the sharing-based channel access protocol, the formula to calculate $P_{success}$ is also the same as the equation in Sec. IV-A, except for (3) and (5), where $\lambda'_{A_d}$ is replaced with $\lambda^{share}_{A_d}$ as defined as follows

$$\lambda^{share}_{A_d} = \lambda'_{A_d}\left(1 - \lambda^{shared}_{STA}\right) + \lambda^{A_d}, \quad (9)$$

where $\lambda^{shared}_{STA}$ is the set of non-AP STAs participating in a shared TxOP, which could be considered as one group with the sharing feature. This implies that as long as any sharing STA obtains the channel, the sharing group would also obtain intrinsically the channel access as well.

*D. Numerical Results*

In order to compare the three protocols described in Sec. III, $\tau = 5\ ms$, $r = 10\ m$, and $\lambda_A = \lambda_B = 0.1$. Furthermore, in order to consider a scenario where the inferences from the OBSS are dominant, $\alpha = 500$ and $\beta = 1$. Moreover, to simulate a case where the trigger frame generation rate is higher than the general frame generation rate from the non-AP STA in the trigger-based scenario, $\omega = 100$.

Fig. 3 shows the success probability of channel access as function of the distance across BSSs. This figure is obtained with $\lambda^{shared}_{STA} = \lambda^{triggered}_{STA} = 0.5$, and highlights that the farther apart the adjacent BSSs are, the easier is for a non-AP STAs in the intended BSS to obtain a channel access.

Fig. 4 shows the success probability of channel access as function of $\lambda^{shared}_{STA}$ or $\lambda^{triggered}_{STA}$. This figure is obtained with $\lambda^{shared}_{STA} = \lambda^{triggered}_{STA}$ and $d = 15m$, and highlights that the more non-AP STAs join the shared TxOP, the less congested the system would be and the easier would be for this group of non-AP STAs to access the channel.

It is also important to note that both Fig. 3 and Fig. 4 highligh the merit of the sharing-based channel access protocol over both the trigger-based and EDCA-based protocol, where the latter is the one achieving the worse performance.

## V. SIMULATION RESULTS

In this section, results obtained through comprehensive system level simulations are provided with the aim to evaluate the performance of the proposed protocol and compare it with the other two state-of-art protocols described in Sec. III. The simulations are performed under the assumptions summarized in Table I. In particular, the simulations are obtained assuming the network topology is composed by 2 BSS, where one is the reference BSS, BSS1, and the other is an OBSS, BSS2. In BSS1, AP1 has four associated non-AP STAs, which are distributed $5m$ away from AP1 in both the horizontal and vertical direction. In the OBSS, AP2 is located $15m$ away from AP1. AP2 also has four associated non-AP STAs, which are distributed $5m$ away from AP2 in both the horizontal and vertical direction.

TABLE I: Main Simulation Parameters

| Parameters | Values |
|---|---|
| Carrier Frequency | 5 GHz |
| Bandwidth | 80 MHz |
| Maximum TxOP | 5ms |
| AP Transmission Power | 21 dBm |
| STA Transmission Power | 15 dBm |
| MCS | 64 QAM w/ 3/4 Code Rate |
| Payload Size | 1400 bytes |

*A. OBSS With Lower Priority Traffic*

Using the aforementioned topology, one of the use cases considered is to study the effects from an OBSS when this is serving lower priority traffic. More specifically, in this case the traffic in each BSS is set as follows:
- In BSS1, each non-AP STA generates one packet every $5ms$ in a constant manner. In case of non trigger-based scenarios, the non-AP STAs contend the channel with the maximum AC (i.e., AC3). On the other hand, in case of trigger-based scenarios, the non-AP STAs transmit UL packets after receiving the trigger frame from AP1, where AP1 sends the trigger frame every $5ms$.
- In BSS2, each non-AP STA is in full buffer mode. The AP2 and its associated non-AP STAs contend the channel with the lowest AC (i.e., AC0). Different traffic conditions are simulated to increase the interference from the OBSS by increasing the number of non-AP STAs associated with OBSS, and the following scenarios are considered:
  – OBSS-L(ight): only AP2 and one of its STAs are actively transmitting;
  – OBSS-M(edium): only AP2 and two of its STAs are actively transmitting;
  – OBSS-L(arge): AP2 and all 4 associated STAs are actively transmitting.

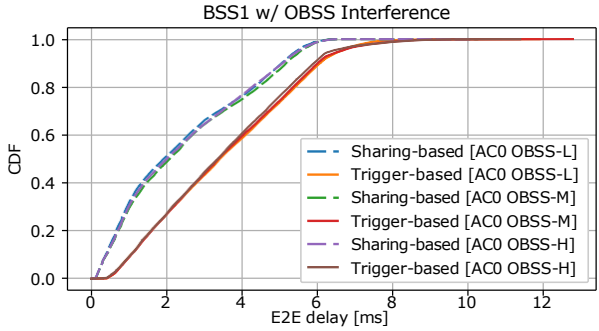

Fig. 5: Cumulative Distribution Function (CDF) of the E2E delay for both the sharing-based and the trigger-based protocol with AC0.

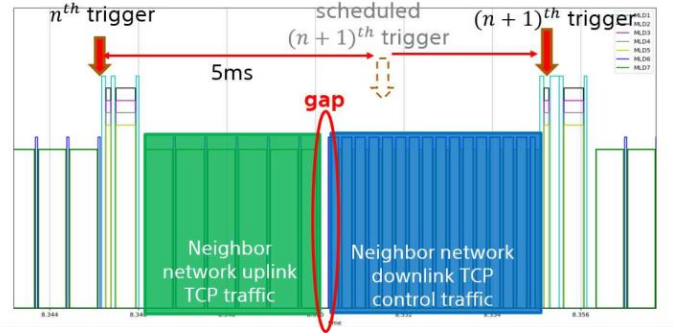

Fig. 6: Illustration of the limitations of the trigger-based protocol.

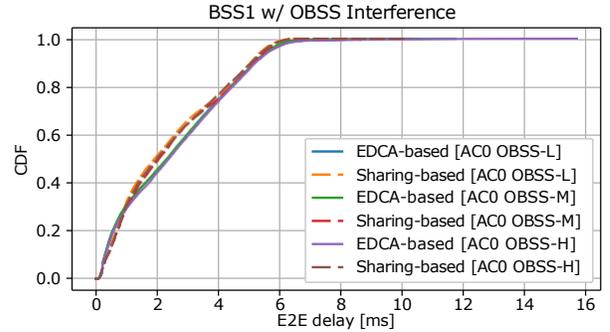

Fig. 7: CDF of the E2E delay for both the sharing-based and the EDCA-based protocol with AC0.

Notice that in this case, as the OBSS has lower AC than the reference BSS, then the contention from the OBSS AP and non-AP STAs has minor impact in the system performance in terms of E2E delay for the non-AP STAs associated with the reference BSS, but what is more relevant is the impact in terms of blockage caused by the existing transmission(s) from the OBSS.

Fig. 5 shows the CDF of the E2E delay of the non-AP STAs in BSS1 for both the trigger-based and the sharing-based protocol with different interference levels from the OBSS. This figure highlights that non-AP STAs in BSS1 experience lower E2E delays when the sharing-based protocol is used compared to the trigger-based protocol, which is due to the nature of the trigger-based protocol, which solicits UL transmissions with a trigger frame, and thus, reduces the channel access flexibility from the perspective of the non-AP STAs. In this matter, Fig. 6 illustrates representative frame transmissions between two consecutive trigger frames. The figure illustrates the presence of a short time gap, i.e., free medium, between two TxOPs hold by the OBSS AP or non-AP STAs. A TxOP for the OBSS UL Transmission Control Protocol (TCP) traffic is depicted in the figure by a green box, and a TxOP for the OBSS Downlink (DL) TCP traffic is instead depicted by a blue box. The trigger frame from BSS1, denoted as $n^{th}$ trigger, initiates a TxOP before the TxOP of the OBSS UL TCP traffic and before a subsequent trigger frame, denoted as $(n + 1)^{th}$ trigger. When the $(n + 1)^{th}$ trigger is generated, the channel is occupied by the TxOP of the OBSS DL TCP traffic, thus, the $(n + 1)^{th}$ trigger cannot be transmitted until after the OBSS finishes its TxOP and nor AP1 is able to obtain access of the channel. In this case, while a short time gap exists between TxOPs hold by the OBSS which could have been used by the reference BSS to acquire the channel and initiate a new TxOP, the $(n + 1)^{th}$ trigger should be postponed resulting in additional queuing delays for the buffered packets in the non-AP STAs of the reference BSS. This is in contrary of what occurs with the EDCA-based or the sharing-based protocol, where the non-AP STAs associated with the reference BSS may hardly miss any gap and can content the channel as needed.

Fig. 7 shows the CDF of the E2E delay for both the EDCA-based and the sharing-based protocol with different interference levels from the OBSS. This figure highlights that that non-AP STAs in BSS1 experience comparable E2E delay with slight improvement when the sharing-based protocol is used compared with the EDCA-based protocol as both are able to content the channel aggressively while the sharing-based protocol is able to help mitigate inferences across all the non-AP STAs.

*B. OBSS With Same Priority Traffic*

Another use case considered in the simulation campaign is to study the effects from an OBSS when this is serving with the same priority traffic as that of the reference BSS. In this case, both the reference BSS and the OBSS use AC3 for channel access, while the rest of the assumptions are the same as that in Sec. V-A. Notice that in this case, as the OBSS has the same AC as the reference BSS, then both the contention from the OBSS AP and its non-AP STAs and the blockage caused by the existing transmission(s) from the OBSS have equal impact on the worse-case E2E delay of the non-AP STAs associated with the reference BSS.

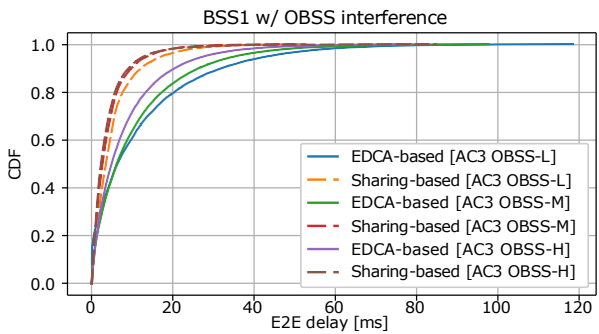

Fig. 9: CDF of the E2E delay for both the sharing-based and the EDCA-based protocol with AC3.

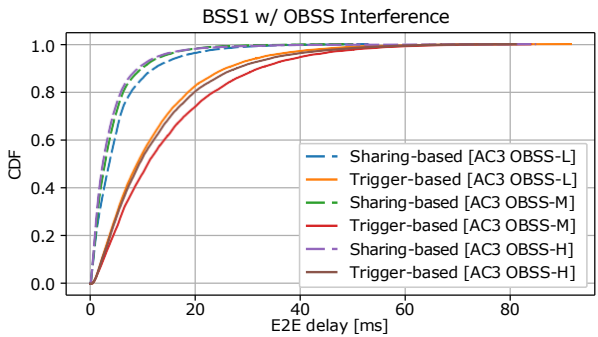

Fig. 8: CDF of the E2E delay for both the sharing-based and the trigger-based protocol with AC3.

Fig. 8 shows the CDF of the E2E delay of the non-AP STAs in BSS1 for both the trigger-based and sharing-based protocol with different interference levels. This figure highlights that the non-AP STAs associated with the reference BSS experience less E2E delay when the sharing-based protocol is used compared with the trigger-based protocol, which again is due to the higher flexibility that the sharing-based protocol is able to offer. Furthermore, this figure shows higher E2E delays than similar results in Sec. V-A, which is due to the increased contention caused by using AC3 at the OBSS.

Finally, Fig. 9 shows the CDF of the E2E delay of the non-AP STAs in BSS1 for both the EDCA-based and sharing-based protocol with different interference levels. This figure highlights that the non-AP STAs in BSS1 experience less E2E delay when the sharing-based protocol is used to compare with the EDCA-based protocol. While with both protocols, any time gap is more aggressively used to contend the channel, once again the sharing-based protocol offers an inherent advantage in reducing not only the intra-BSS contention but also the inter-BSS contention.

## VI. CONCLUSION

In this paper, to cope with the stringent requirements that next-generation of WLANs require, a new channel access procedure is proposed, which allows to further mitigate the channel access contention in such systems by allowing co-operation among STAs while maintaining high channel access flexibility compared to other access procedures, which are currently adopted by the Wi-Fi 802.11 standard. After a detailed description of the proposed procedure and comparison in terms of conventional transmission timeline with the two state-of-art protocols, i.e., EDCA-based and trigger-based protocol, a statistical model is provided and closed form expressions to compare all three protocols in terms of the success probability of channel access for the non-AP STAs within the coverage area of a reference BSS in presence of an OBSS are derived. Furthermore, a comprehensive system level analysis of the three protocols is provided. Both analytical and system level analysis show that the proposed protocol outperforms the two state-of-art protocols in terms of both mitigating the E2E delay and allowing a better spectrum utilization by reducing the overall contention in system.